\documentclass[prd,twocolumn]{revtex4}
\usepackage{graphicx, epsfig}
\usepackage{color}
\usepackage{mathrsfs}
\usepackage{amsmath}

\newcommand{\be}{\begin{equation}}
\newcommand{\ee}{\end{equation}}
\newcommand{\bea}{\begin{eqnarray}}
\newcommand{\eea}{\end{eqnarray}}

\newcommand{\gapp}{\mathrel{\raise.3ex\hbox{$>$}\mkern-14mu
              \lower0.6ex\hbox{$\sim$}}}
\newcommand{\lapp}{\mathrel{\raise.3ex\hbox{$<$}\mkern-14mu
              \lower0.6ex\hbox{$\sim$}}}

\begin{document}
\title{Can primordial magnetic fields seeded by electroweak strings cause an alignment of quasar axes on cosmological scales?}
\author{Robert Poltis}
\author{Dejan Stojkovic}
\affiliation{HEPCOS, Department of Physics,
SUNY at Buffalo, Buffalo, NY 14260-1500}
 %%%%%%%%%%%%%%%%%%%%%%%%%%%%%%%%%%%%%%%%%%%%%%%%%%%%%%%

\begin{abstract}
The decay of non-topological electroweak strings formed during the electroweak phase transition in the early universe may leave an observable imprint in the universe today. Such strings can naturally seed primordial magnetic fields. Protogalaxies then tend to form with their axis of rotation parallel to the external magnetic field, and moreover, the external magnetic field produces torque which forces the galaxy axis to align with the magnetic field, even if the two axis were not aligned initially.
This can explain an (observed, but as of yet unexplained) alignment of the quasars' polarization vectors. We demonstrate that the shape of a magnetic field left over from two looped electroweak strings can explain the non-trivial alignment of quasar polarization vectors and make predictions for future observations.

\end{abstract}

\maketitle

\section{Introduction}

Recently, Hutsem\'{e}kers \cite{Hutsemekers} made two interesting observations on a sample of 355 quasars\footnote{In this paper we use the term 'quasar' to describe both optically and radio selected quasi-stellar objects.}. They observed that the polarization vector of quasars appear to be i.) somewhat aligned over large (cosmologically interesting) volumes of space and ii.) the angle of these vectors seem to rotate coherently with increasing redshift. As discussed in their paper, these two observations seem unlikely to be attributable to either natural contamination such as intervening dust particles or unaccounted instrumental bias. Instead, the effect appears to be cosmological.

The direction of the optical polarization vector can be attributed to the physical orientation of the quasar itself \cite{Elvis,BorguetQuasarPolarizationAxisDirection,BorguetSmallPaper}. We propose that the quasars themselves are somewhat aligned on cosmological scales. Any model that explains the coherent alignment of quasars on such scales should also address the rotation through $\sim 240^{o}$ that is observed in the sample. This feature cannot be easily accommodated in generic models.

We propose that the orientation of these quasars is caused by a magnetic field left over from two linked loops of electroweak strings. From the time of the electroweak phase transition to today, magnetic field lines seeded by these strings are stretched by the expansion of the universe and act as a background magnetic field at the time of quasar formation.
We fit the alignment data and find that the electroweak string loops can explain this alignment very well. We emphasize that our explanation is based on known and pretty well understood physics of the standard model and its embedded defects like electroweak strings.

\section{The Data}

The observation of the quasar polarization vectors was carried out at the European Southern Observatory in Chile from August 2000 through October 2003. The quasars themselves are all located at high galactic latitudes ($|b| \geq 30^{o} $) towards both the north and south galactic poles. Objects that were given preference for observation were bright quasars, as well Broad Absorption Line, radio-loud and red quasars. In addition, based on findings from \cite{HutsemekersPaper1} and \cite{HutsemekersPaper2}, special emphasis was given to objects in two regions where an alignment effect was previously observed. One of these regions lies towards the north galactic pole and is delimited in right ascension and redshift by $11^h15^m \leq \alpha \leq 14^h29^m$ and $1.0 \leq z \leq 2.3$. The other region lies in the direction of the south galactic pole and is delimited by $21^h20^m \leq \alpha \leq 24^h00^m$ and $0.7 \leq z \leq 1.5$. The median polarization of the 355 quasars is approximately $1.38\%$ with no object having a polarization less than $0.6\%$. Every object observed possessed an uncertainty in the polarization angle of no more than $14^{o}$. With one exception, no object studied possesses a redshift greater than $3$.

\section{The alignment effect}

\begin{figure}[ht]
\includegraphics[width=3.4in]{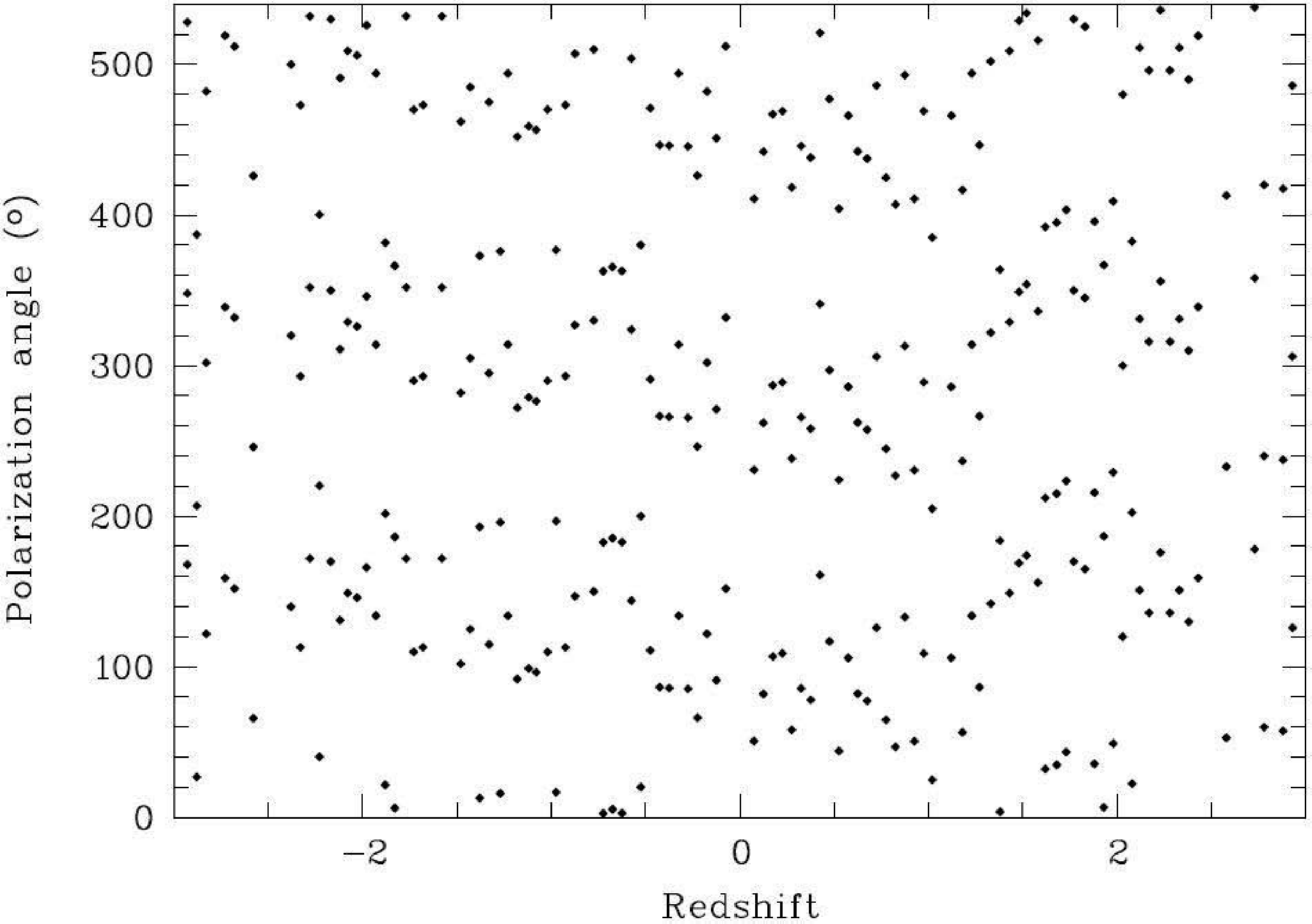}
\caption{The polarization angle of all 355 quasars. All polarization angles are vectorally averaged and placed in redshift bins of $\Delta z=0.05$. Following the convention of \cite{Hutsemekers}, redshifts in the direction of the North Galactic Pole (NGP) are counted positive and redshifts in direction of the South Galactic Pole (SGP) are counted negative. To help identify a pattern, each polarization angle is plotted three times: at a point $(z,\theta)$, $(z,\theta +180^o)$, and $(z,\theta +360^o)$. (Figure taken directly from \cite{Hutsemekers}.)}
\label{355quasars}
\end{figure}

The polarization vectors appear to be coherently aligned over large volumes of space with a probability of less than $0.1 \%$ of such an alignment occurring by chance \cite{Hutsemekers}. In addition, the direction of this alignment also appears to rotate with redshift. All 355 objects sampled are included in Fig.~\ref{355quasars}. From Fig.~\ref{355quasars} there is an apparent relation between polarization angle and the redshift of the source.

\section{Quasar structure}
\label{Quasar Structure}
\begin{figure}[h]
\includegraphics[width=3.4in]{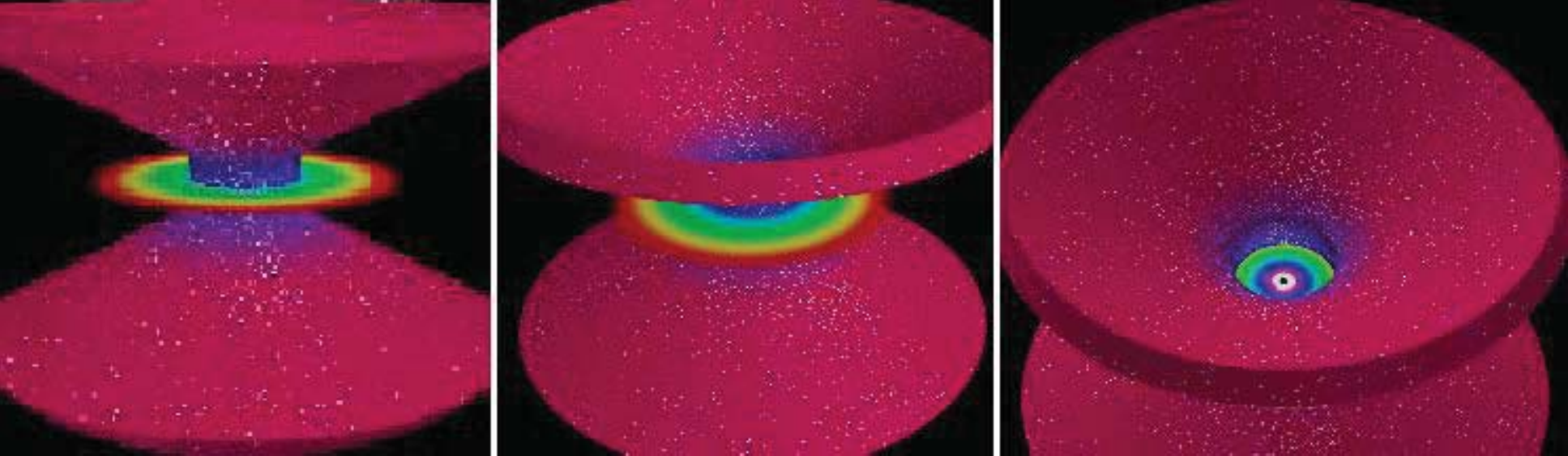}
\caption{Three views of the inner regions of a quasar. Left: a view from the side, nearly in the plane of the accretion disk. Center: a view down the outflow, along the funnel-shaped winds emanating from the accretion disk. The quasar in this center picture would be observed as a BAL quasar. Right: a view from above the funnel-shaped outflow. From this last perspective, there is an unobstructed view of the continuum source. (Figures taken directly from \cite{Elvis}.)}
\label{Quasar3Views}
\end{figure}

\begin{figure}[ht]
\includegraphics[width=3.6in]{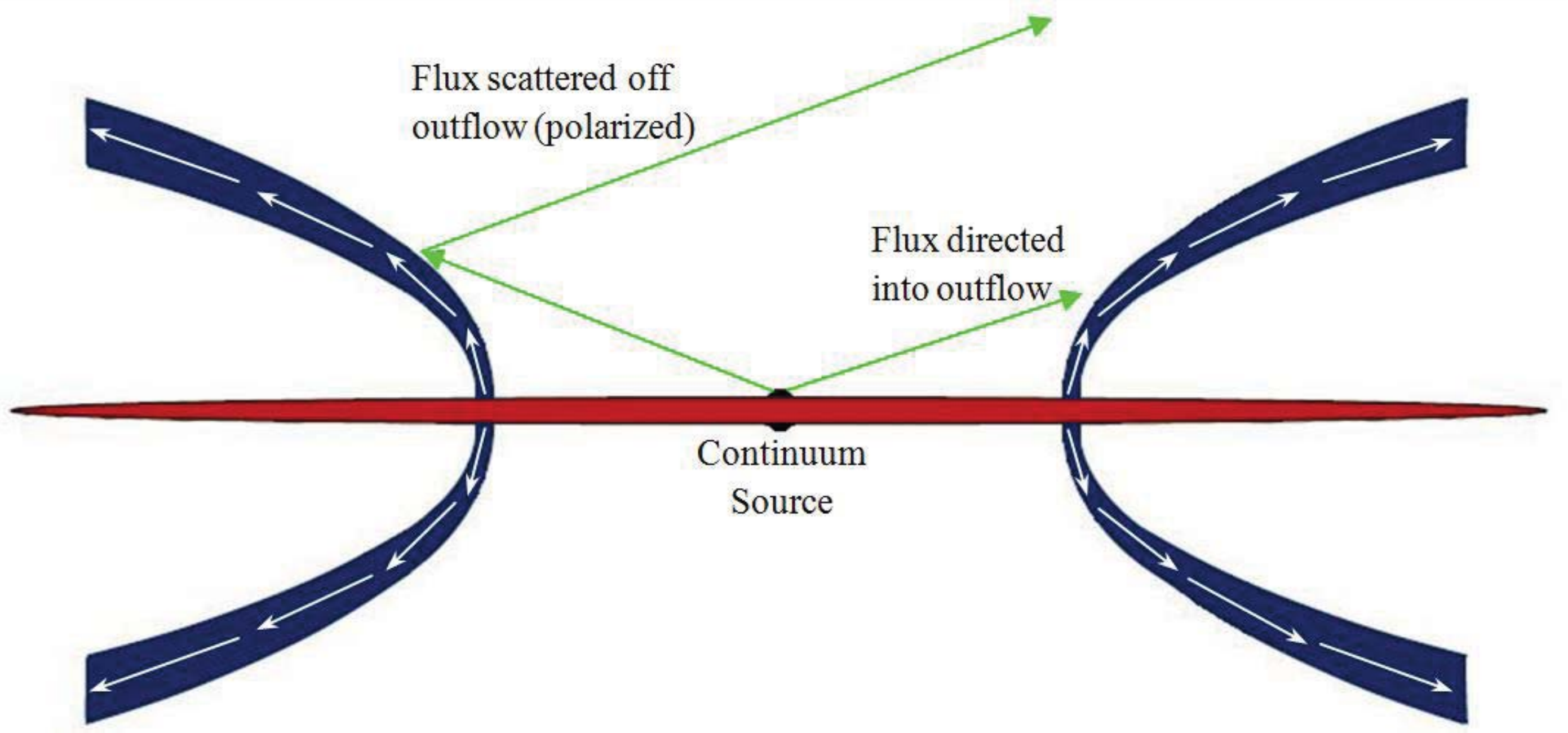}
\caption{A cross-section of the inner regions of a quasar as described in \cite{Elvis}. An accretion disk surrounds a continuum source at the center of the object. The direction of the outflow winds is shown by the white arrows. Observers looking directly into the conical outflow of the object will observe a BAL quasar. Flux emanating from the continuum source will be Thompson scattered off the outflow and polarized.}
\label{QuasarCrossSection}
\end{figure}

At the center of a quasar lies a super-massive black hole surrounded by an accretion disk. The central regions of quasars emit massive amounts of continuum radiation. From this accretion disk, a warm wind arises perpendicular to the plane of the accretion disk over a narrow range of radii. Radiation pressure then accelerates this wind radially away from the continuum source, causing a funnel shape outflow (Figs.~\ref{Quasar3Views},~\ref{QuasarCrossSection})\cite{Elvis}.

A subclass of quasars, Broad Absorption Line (BAL) quasars comprise $\sim10\%$ of all quasars. These objects (which were given observational preference in \cite{Hutsemekers}) are observed as such because of their orientation to us \cite{Elvis}. The polarized radiation from BAL quasars originates from the conical shell surrounding the center of the quasar \cite{Elvis}. Flux emanating from the continuum source at the center of the quasar is Thompson scattered off the shell (Fig.~\ref{QuasarCrossSection}). This scattering off the flow explains the observed $\sim10\%$ polarized continuum observed in troughs compared to the $\sim0.5\%$ polarization observed in non-BAL quasars\cite{Elvis}.

In \cite{BorguetQuasarPolarizationAxisDirection,BorguetSmallPaper} the authors report a correlation between the major axis of the host galaxy of a quasar and the direction of linear polarization of the object. The direction of polarization from Type 1 quasars (quasars whose spectrum contain broad emission lines) tends to be parallel with the axis of the quasar host while the direction of polarization from Type 2 quasars (quasars whose spectrum contains only narrow emission lines) tends to be perpendicular to the host axis. Quasars that have broad emission line spectra will on average also have broader absorption lines as well \cite{TrumpBALBEL}, and the authors of \cite{Hutsemekers} gave preference to BAL quasars in their observations.

\section{Electroweak cosmic strings in the early universe}

The early universe very likely went through a number of phase transitions that gave rise to various topological defects via the Kibble Mechanism. Gauge groups of grand unifying theories are complicated enough to give rise to magnetic monopoles, domain walls and cosmic strings. While the electroweak standard model gauge group $SU(2)_L \times U(1)_Y$ does not contain non-trivial topology, it does contain so-called embedded defects, most notably electroweak cosmic strings \cite{Nambu:1977ag,Vachaspati:1992jk,Barriola:1993fy}. In the minimal version of the standard model, for the physical values of the relevant parameters (weak mixing angle and Higgs mass), electroweak cosmic strings are not stable configurations \cite{James:1992wb,Stojkovic:2000ix,Starkman:2001tc}. Thus, it is very unlikely that they can survive till today. However, one can not avoid their formation and a subsequent decay. It would be very interesting if one could find an imprint left by the electroweak strings that can be observed today. The most promising effects would perhaps be associated with primordial magnetic fields seeded by the electroweak strings.

\begin{figure}[ht]
\includegraphics[width=3.4in]{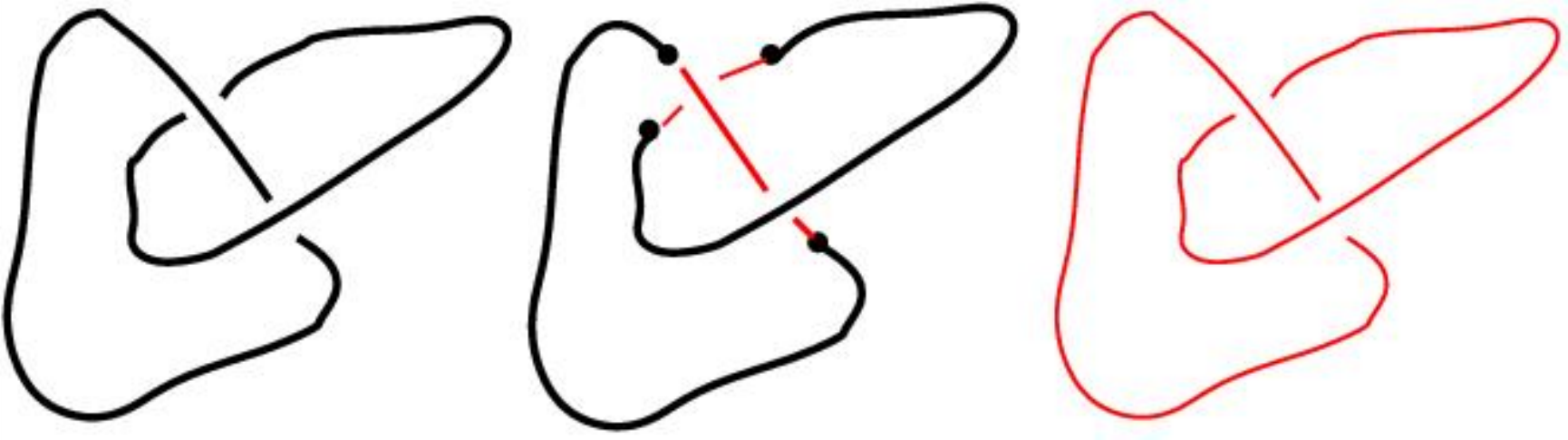}
\caption{The picture on the left shows two linked strings. In the middle picture, a monopole-anti-monopole pair is created. The strings, now broken, will collapse and leave behind a remnant magnetic ($\vec{B}$) field (right picture). (Figure taken directly from \cite{VachaspatiLinkedStrings}.)}
\label{LinkedStrings}
\end{figure}

Cosmic strings produced during the electroweak phase transition can not be open-ended-that is, they will exist as either closed loops or be infinitely long\footnote{Cosmic strings may also exist in more interesting ways such as strings that terminate on magnetic monopoles}\cite{VachaspatiEWStringsReview}. Cosmic strings also contain small scale structure in the form of wiggles \cite{VachaspatiWigglyStrings,VachaspatiVilenkinWigglyStrings,MartinVilenkinTopDefects}. Now consider that case of two electroweak strings that are initially linked (as shown in the left picture of Fig.~\ref{LinkedStrings}). The two linked strings each carry Z lines of magnetic flux. The direction of magnetic flux is parallel to the direction of the string \cite{VachaspatiEWStringsReview}.

\begin{figure}[ht]
\includegraphics[width=3.4in]{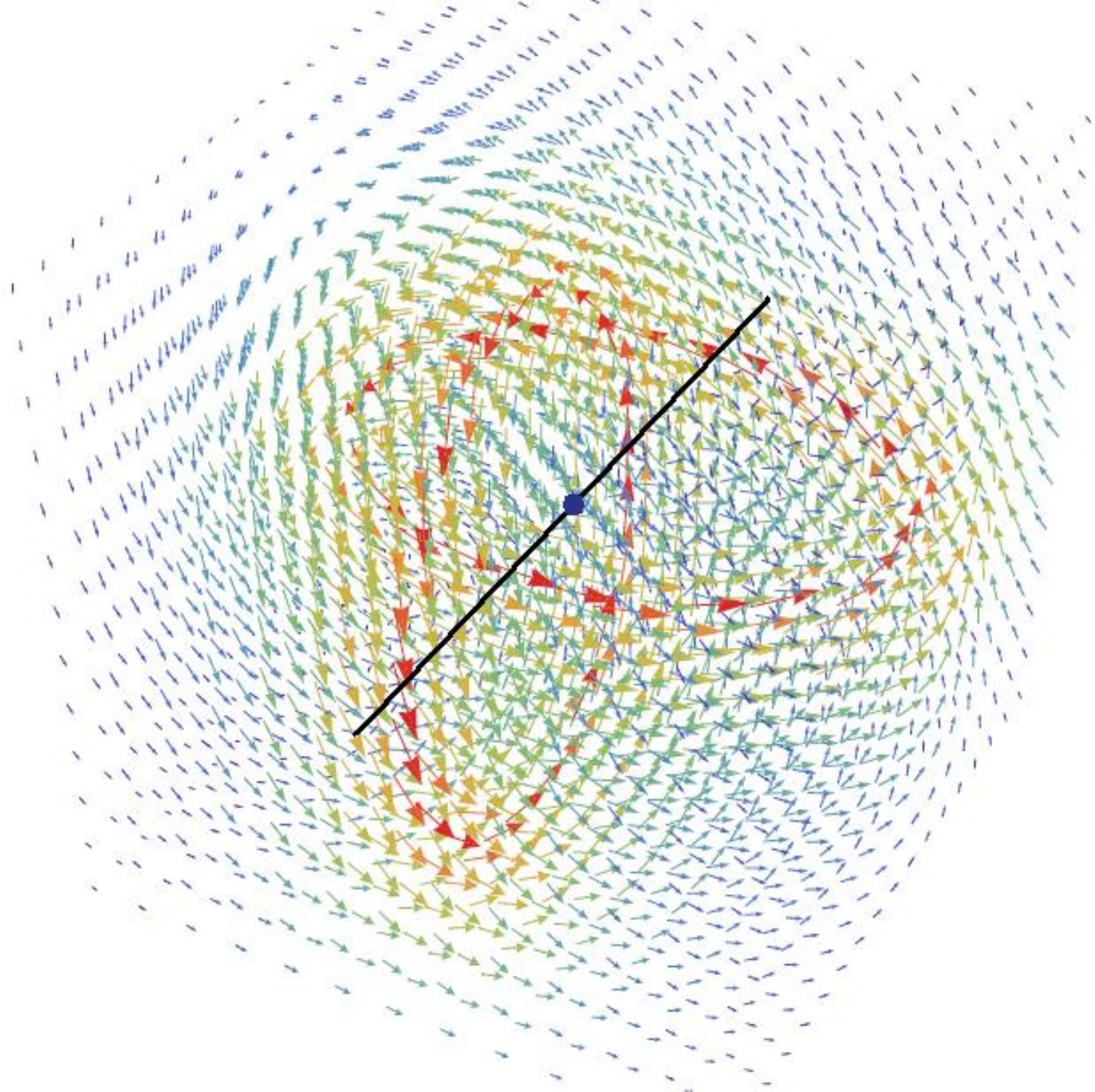}
\caption{The shape of the magnetic field that would exist today as a result of the decay of two linked strings in the early universe, as described by Eq.~(\ref{TwoLoops}). Drawn in are the A1-A3 axis (black line) and location of Earth (blue dot) as described in sec. (\ref{MatchingModelObservation}). This figure is intended to give the reader an idea of the shape of the magnetic field that we are discussing. Each loop today has a radius on the scale of Gpc.}
\label{LinkedBField}
\end{figure}

A mechanism that may cause strings to decay is through the creation of a monopole-anti-monopole pair. Once the monopole-anti-monopole pair is created, tension in the string will pull them apart. The Z magnetic flux lines will become frozen into the highly conductive plasma of the early universe. After these strings have been destroyed, a linked magnetic ($\vec{B}$) field will remain \cite{VachaspatiLinkedStrings,VachaspatiEWStringsReview}. This magnetic field can than be carried by the expansion of the universe. From the time that these linked strings decay around the electroweak phase to today, the left-over magnetic field configuration would be carried by the expansion of the universe and today exist on cosmological scales. Requiring $\nabla\cdot\vec{B}=0$ in an Abelian theory implies that parallel magnetic field lines will repel \cite{VachaspatiEWStringsReview}. Therefore, our final field configuration should appear as two spread out, interconnected loops of magnetic field (Fig.~\ref{LinkedBField}).

\section{Form of the resulting B-field}

As mentioned before, the magnetic flux from a Z-string will lie along the length of the string. For simplicity, we will assume the shape of the magnetic field left over from an electroweak string to be a circle.

For a single magnetic field loop of radius $R$ in the $z=0$ plane in cylindrical coordinates, we want the magnetic field to reach a constant value ($B_o$) around the string and go to zero infinitely far away from the string, $\rho\rightarrow\infty$ and $z\rightarrow\pm\infty$. Such a magnetic field can be expressed by
\be
\vec{B}=\frac{B_o}{R}\exp\left[{-\sqrt{z^2+(\rho-R)^2}}\right]\hat{\phi}
\label{OneLoop}
\ee

From the form of Eq.~(\ref{OneLoop}), it is obvious that such a field satisfies the Maxwell Equation $\nabla\cdot\vec{B}=0$.

The easiest way to describe the magnetic field resulting from the decay of two linked strings is to change to cartesian coordinates and place one loop of magnetic field (of radius $R$) in the $z=0$ plane centered a distance $d$ from the origin, where $0<d<2R$, and place a second loop (for simplicity, also of radius $R$) lying in the $x=0$ plane centered at the origin. Assuming the magnetic field from both strings has the same magnitude ($B_o$) such a magnetic field configuration will be a vector sum of the two fields.
\begin{widetext}
\bea
\vec{B}\!\!\!\!\!&&=\frac{B_o}{R}\left[\left(-(y+d)\exp\left[-\sqrt{z^2+\left(\sqrt{x^2+\left(y+d\right)^2}-R\right)^2}\right]\right)\hat{x}\right. \nonumber \\
&&+\left(x\exp\left[-\sqrt{z^2+\left(\sqrt{x^2+\left(y+d\right)^2}-R\right)^2}\right]-z\exp\left[-\sqrt{x^2+\left(\sqrt{y^2+z^2}-R\right)^2}\right]\right)\hat{y} \nonumber \\
&&+\left.\left(y\exp\left[-\sqrt{x^2+\left(\sqrt{y^2+z^2}-R\right)^2}\right]\right)\hat{z}\right]
\label{TwoLoops}
\eea
\end{widetext}
Eq.~(\ref{TwoLoops}) describes two perpendicular loops of magnetic fields with amplitude $B_o$ and radius $R$, as shown in Fig.~\ref{LinkedBField}. It is also possible to describe two looped magnetic fields that are not exactly perpendicular to each other.  To do so, we can rotate the string in the $x=0$ plane by an angle $\psi$ about the y-axis. This modifies the form of the loop originally in the $x=0$. The rotated loop of magnetic field is given by Eq.~(\ref{AngledLoop}).
\begin{widetext}
\vspace{-0.2cm}
\bea
\vec{B}=&&\!\!\!\!\!\frac{B_o}{R}\exp\left[-\sqrt{\left(x\cos\psi-z\sin\psi\right)^2 + \left(\sqrt{y^2+\left(x\sin\psi+z\cos\psi\right)^2}-R\right)^2}\right] \nonumber\\
&&\left[y\sin\hat{x}-\left(x\sin\psi+z\cos\psi\right)\hat{y}+y\cos\psi\hat{z}\right]
\label{AngledLoop}
\eea
\end{widetext}
In our analysis, we found the data to best match the case where the two magnetic fields are perpendicular to each other; the case where $\psi=0$.

\section{Effects of magnetic flux on quasar alignment}
\label{EffectFluxAlignment}

In sec. (\ref{Quasar Structure}) we presented a correlation between the angle of polarization and the orientation of the quasar host. To explain the correlation between a background magnetic field and quasar host alignment we will assume a matter dominated universe ($a\propto t^{2/3}$) from the present time back to the time of recombination and a radiation dominated universe ($a\propto t^{1/2}$) from the time of recombination back to the electroweak phase transition. Assuming that the electroweak phase transition occurs at $t\sim 10^{-11}s$ and the time of recombination to be $t\sim 10^{13}s$, we find that the electroweak phase transition occurs at a redshift of $z\sim 10^{12}$.

The data collected by \cite{Hutsemekers} extends out to redshifts $z\sim 3$, implying that the looped magnetic fields are on the scale of Gpc. Taking a single magnetic field loop to be of radius $R_o\sim 1$ Gpc today leads to the magnetic field loop to be $R_{rec}\sim 1$ Mpc at the time of recombination (assuming $z_{rec}\sim 10^{3}$). Working back further, we find that the loop at the electroweak phase transition (at $z\sim 10^{12}$) to be of size $R_{EW}\sim 10^{-6}$pc, or $R_{EW}\sim 10^{10}$m (again, here we are only making an order of magnitude estimate). Although this scale is larger than the causally connected universe at the time of the electroweak phase transition, there is no reason that such a string should be unphysical, since extended topological defects naturally have superhorizon structures.

We should note that the strings themselves may have small structural irregularities, i.e. wiggles. The wiggles on cosmic strings exist down to a scale $\ell\sim\alpha t$. The parameter $\alpha$ is not known, but simulations suggest that at most $\alpha \lesssim 10^{-3}$ \cite{MartinVilenkinTopDefects}. Taking $\alpha\sim 10^{-3}$, the wiggles on an electroweak sting will exist down to a characteristic length scale $\ell\sim 3\times 10^{-6}$m at the time of the string formation. Stretched by the expansion of the universe, this scale today would be of the order of $\ell\sim 10^{-7}$pc, several orders of magnitude below the scale of quasars. At the time of galaxy and/or quasar formation, the magnetic field would also contain wiggles that may be relevant for seeding the magnetic field and Eq.~(\ref{EulerEq}).

It is well known that a magnetic field, with the present-day strength of about $10^{-9}$Gauss and more or less coherent structure on Mpc or larger
scales, can strongly influence early galaxy formation (for a review see \cite{Grasso:2000wj} and references therein). The primordial magnetic fields alone likely cannot 
be responsible for the observed galaxy power spectrum on large scales. However, it seems quite likely that magnetic fields do play a significant role by introducing a bias on the
formation of galaxy sized objects \cite{Kim}. It is also argued in  \cite{Coles} that somewhat inhomogeneous
magnetic field could modulate galaxy formation in the cold dark matter picture by giving the baryons a streaming velocity relative to
the dark matter.

The magnetohydrodynamic equations governing the evolution of linear density perturbations are \cite{Grasso:2000wj,Kim}

\bea
\rho\!\left(\frac{\partial \vec{v}}{\partial t}\!+\!\frac{\dot{a}}{a}\vec{v}\!+\!\frac{\left(\vec{v}\cdot\nabla\right)\vec{v}}{a}\right)\!=\!-\frac{\nabla p}{a}\!-\!\rho\frac{\nabla\phi}{a}\!-\!\frac{\left(\!\nabla\!\times\!\vec{B}\!\right)\!\times\!\vec{B}}{4\pi a} 
\label{EulerEq}
\eea
\bea
\frac{\partial \rho}{\partial t}+3\frac{\dot{a}}{a}\rho +\frac{\nabla \cdot \left(\rho\vec{v}\right)}{a}=0
\label{ContinuityEq}
\eea
\bea
\frac{\nabla ^2\phi}{a^2}=4\pi G\left[\rho-\rho_b\left(t\right)\right]
\label{PoissonsEq}
\eea
\bea
\nabla\cdot\vec{B}=0
\label{GaussLawForMag}
\eea
\bea
\frac{\partial}{\partial t}\left(a^2\vec{B}\right)=\frac{\nabla\times\left(\vec{v}\times a^2\vec{B}\right)}{a}
\label{FaradaysLaw}
\eea

The Lorentz force perturbs the smooth background density $\rho _b(t)$ inducing density perturbations $\delta\rho(\vec{x},t)$ and peculiar velocities $\vec{v}(\vec{x},t)$ within the fluid. The peculiar velocity of the baryonic fluid will in turn backreact to create an additional magnetic field $\delta\vec{B}(\vec{x},t)$. Following the notation of \cite{Kim}, we introduce the small quantity $\delta$ where 
\bea
\rho\left(\vec{x} ,t\right)=\rho _b\left( t\right) +\delta\rho\left(\vec{x} ,t\right)\equiv\rho _b\left( t\right) [1+\delta\left(\vec{x} ,t\right) ]
\eea
The total magnetic field is therefore the initial background magnetic field leftover from electroweak strings and the magnetic field resulting from fluid backreaction.
\bea
\vec{B}\left(\vec{x} ,t\right)=\vec{B}_b\left(\vec{x} ,t\right)+\delta\vec{B}\left(\vec{x} ,t\right)
\eea

Assuming that density perturbations, peculiar velocities and induced magnetic fields resulting from the Lorentz force are small, we can linearize eqs.~(\ref{EulerEq}-\ref{FaradaysLaw}) in $\vec{v}$ and $\delta$ which become

\bea
\frac{\partial \vec{v}}{\partial t}+\frac{\dot{a}}{a}\vec{v}=-\frac{\nabla\phi}{a}+\frac{\left(\nabla\times\vec{B}_b\right)\times\vec{B}_b}{4\pi a\rho_b}
\label{LinEulerEq}
\eea
\bea
\frac{\partial\delta}{\partial t}+\frac{\nabla\cdot\vec{v}}{a}=0
\eea
\bea
\nabla ^2\phi =4\pi a^2G\rho_b\delta
\eea
\bea
\nabla\cdot\vec{B}_b=\nabla\cdot\delta\vec{B}=0
\eea
\bea
\frac{\partial}{\partial t}\left( a^2\vec{B}_b\right) =0
\label{LinFaradayBackground}
\eea
\bea
\frac{\partial}{\partial t}\left( a^2\delta\vec{B}\right) =\frac{\nabla\times\left(\vec{v}\times a^2\vec{B}_b\right)}{a}
\eea

Because we are more concerned with general behavoir of the baryonic fluid, we will look at the linearized MHD equations that describe the evolution of $\vec{v}$ to zeroth order in $\delta$ and $\delta\vec{B}$. The linearized Faraday's Law implies that the background magnetic field will evolve as 
\bea
\vec{B}_b\left(\vec{x},t\right) =\vec{B}_b\left(\vec{x}\right)\exp ^{-2Ht}
\label{BackgroundBFieldTimeDependence}
\eea
where $\vec{B}_b$ is the magnetic field at the beginning of structure formation and $H$ is the Hubble constant.

The linearized Euler equation (eq.~(\ref{LinEulerEq})) can be rewritten as
\bea
\frac{\partial\vec{v}}{\partial t}=-\frac{\dot{a}}{a}\vec{v}-\frac{\nabla\phi}{a}+\frac{\mu _o}{4\pi a\rho _b}\vec{j}\times\vec{B}_b
\eea
To determine the correlation between the direction of the magnetic field and quasar orientation, we assume that locally (on the scale of quasar formation) the average magnetic field lies only in the z-direction: $\langle B_{x}\rangle =\langle B_{y}\rangle =0$, $\langle B_{z}\rangle\neq 0$. The magnetic field may contain structure on scales down to the typical wiggle scale $\ell\left(\sim 10^{-7}pc\right)$, but the overall average field must lie along the large scale direction of the string. If we now allow ourselves to write the background magnetic field as $\vec{B}_b(\vec{x},t)=0\hat{x}+0\hat{y}+B_b\exp^{-2Ht}\hat{z}$ where now $B_b$ is assumed constant we obtain
\bea
\frac{\partial\vec{v}}{\partial t}=-\frac{\dot{a}}{a}\vec{v}-\frac{GM}{ar^2}\hat{r}-\frac{\mu _o}{4\pi a}\frac{\rho}{\rho _b}\vert B_b\vert\exp ^{-2Ht}v_{\perp}\hat{\phi}
\label{SimpleStructureFormation}
\eea
where $v_{\perp}$ is the component of peculiar velocity that is perpendicular to the z-direction (string/magnetic field direction). 

The first term in eq. (\ref{SimpleStructureFormation}) acts like a friction term caused by the expansion of the universe. The second term describes the collapse of a non-rotating spherical protogalactic cloud due to gravitational contraction. The final term affects the peculiar velocities of cloud particles via the Lorentz force. An important question to ask is which terms in eq. (\ref{SimpleStructureFormation}) will come to dominate (or at least significantly affect) a collapsing protogalaxy.

To compare the size of the gravitational and magnetic effects on a collapsing protogalaxy, we assume a spherical collapsing protogalaxy of uniform density and Hubble constant $H\sim70kms^{-1}Mpc^{-1}\sim2.3\times10^{-18}s^{-1}$. The smallness of $H$ allows us to ignore the friction-like term caused by the expansion. Defining $\Xi$ as the ratio of the strength of the gravitational term to the Lorentz term in Eq. (\ref{SimpleStructureFormation}) and neglecting the time dependence of the background field strength we find
\bea
\Xi =\frac{\frac{4}{3}\pi G\rho_b r}{10^{-7}\vert B_b\vert v_{\perp}}
\label{RatioGravToMag}
\eea
Assuming a baryon density of $3.8\times 10^{-28}kg m^{-3}$ today and background magnetic field strength $\vert B_b\vert =10^{-12}G$, the ratio $\Xi$ for a protogalactic cloud located around $z=9$ with temperature $T=10K$ will be $\Xi_{H^{+}} =80$ per kpc for $H^+$ ions in the cloud and $\Xi_{e^{-}} =1.9$ per kpc for electrons in the cloud. Near the central regions of the cloud over a length scale of lightyears (comparable to the scale for quasars) this ratio is $\Xi_{H^{+}} =2.46\times 10^{-2}$ per LY and $\Xi_{e^{-}} =5.78\times 10^{-4}$ per LY. This implies that the inner regions of a collapsing cloud (including scales comparable to the size of quasars) will be dominated by effects of the background magnetic field while the outer regions of the collapsing protogalactic cloud will be more affected by gravitational contracting than the Lorentz force.

The angular momentum of a quasar will most likely point in a direction parallel to the average direction of the background magnetic field. Because objects interact with their surroundings from the time of formation, there is no guarantee that an individual quasar will not have changed its orientation since formation, but because every quasar (in a certain cosmologically interesting volume of space) formed in a similar background, we still expect to see some trends in the data. We therefore expect that the average quasar polarization direction be parallel to the direction of the magnetic field.

An alternative mechanism that could compliment the previously mentioned mechanism of alignment of quasar hosts is that the magnetic field physically flips the quasar host itself. A magnetic dipole lying in an external magnetic field will experience a torque ($\tau$) acting on it. Using basic mechanics, we can perform an order of magnitude calculation to determine a typical flipping time for a galaxy given a certain magnetic field ($\vec{B}$) and dipole ($\vec{m}$) strength.

Suppose we describe a quasar as a solid disk with moment of inertia $I$ and magnetic dipole moment $\vec{m}$ in an external magnetic field $\vec{B}$. The work necessary to rotate the quasar about its diameter can be written as

\be
W=\int\tau d\theta =\int mB\sin (\theta )d\theta
\label{Work}
\ee

The maximum possible work done on the quasar would involve rotating the quasar through an angle of $90^o$, so that we may say

\be
W_{max}=mB
\label{MaxWork}
\ee

Work can also be described as a change in kinetic energy, which may be written as

\be
W=\Delta KE=\frac{1}{2}I\Delta \omega ^2
\label{WorkKE}
\ee

Combining Eqs.~(\ref{MaxWork}) and (\ref{WorkKE}) and recalling that torque can also be described by $\tau = I\alpha = I\frac{\Delta \omega}{\Delta t}$, we can see that a typical flip time is given by

\be
\Delta t=\sqrt{\frac{2I}{mB}}
\label{FlipTime}
\ee

Using fiducial values of $10^{11}M_\odot$, $15$kpc, $10^{63}$J/T and $10^{-12}$T for a quasar host's mass, radius, dipole moment, and the strength of the external magnetic field (respectively), we find that the time for a quasar to rotate through an angle of $90^o$ is $\sim 4.6\times10^{15}$s$\sim 1.5\times 10^8$years. This timescale is significantly less than the age of the universe at $z=3$, thereby allowing the quasars in the sample sufficient time to allign their axes with the external magnetic field.

We have presented two effects that work synergetically, galaxies themselves (and their quasars) prefer to form with the rotational axis parallel to the external magnetic field, and
the external magnetic field tends to align the quasar axis with itself even if initially the two axis were not aligned.

\section{Matching the model with observation}
\label{MatchingModelObservation}

\begin{figure}[h]
\includegraphics[width=3.4in]{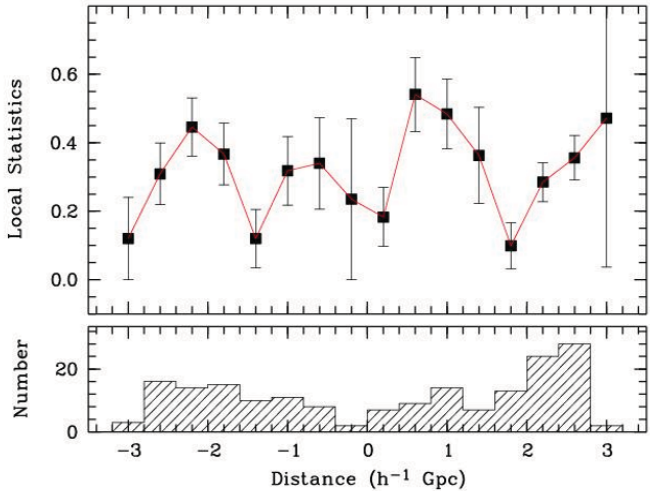}
\caption{A plot of the local statistics along the A1-A3 axis as described in \cite{Hutsemekers}. A high local statistic value indicates a high degree of alignment of quasar polarization vectors. Bins of size $\Delta r=0.4h^{-1}$ Gpc are used. Instead of redshift, comoving distances of $r=6\left(1-\left(1+z\right)^{-1/2}\right)h^{-1}$ Gpc are used where $h$ is the Hubble constant in units of 100 kms\textsuperscript{-1}Mpc\textsuperscript{-1}. Again, objects in the direction of the NGP are assigned positive distances and objects in the direction of the SGP are assigned negative distances. The histogram below the graph gives the number of quasars in each bin. Figure taken directly from \cite{Hutsemekers}.}
\label{A1A3LocalStats}
\end{figure}

A pattern of alignment is apparent in Fig.~\ref{355quasars}. The authors of \cite{Hutsemekers} notice an especially high degree of alignment within two regions of space referred to as the A1-A3 axis. This region of space contains 183 of the 355 observed quasars. They plot the average angle of the quasar polarization vectors in bins of $\Delta z=0.5$ and propose a linear best fit for the rotation of the quasar polarization vectors given by $\bar\theta =268-42z$. This constitutes a rotation through $\sim250^o$ from $z=-3$ through $z=3$. Within the A1-A3 axis the authors of \cite{Hutsemekers} also observe a varying degree of alignment among the quasars as compared to other nearby quasars.  This varying degree of alignment is shown in Fig.~\ref{A1A3LocalStats}.

Our model allows a full rotation through $\sim270^o$ to match with observation. Specifically, if we lie near the center of two linked strings, it would be possible to observe a rotation through $\sim135^o$ as we look out in opposite directions towards the NGP and SGP. Suppose that the A1-A3 axis coincides with the $x=z=0$ line of Eq.~(\ref{TwoLoops}). The magnetic field along that line would take the form of Eq.~(\ref{BAlongYAxis}). For generality, we also included a term $s$ to allow the two loops to be shifted along the y-axis. Therefore, in a plot of $\vec{B}(x,y,z)$, we can place the Earth at the origin.

\begin{eqnarray}
\vec{B}=&&
-\frac{B_o}{R}((y+s)+d)\exp\left[-\lvert \lvert (y+s)+d \rvert -R \rvert \right]\hat{x} \nonumber \\ && +\frac{B_o}{R}(y+s)\exp\left[-\lvert \lvert y+s \rvert -R \rvert \right]\hat{z}
\label{BAlongYAxis}
\end{eqnarray}

\begin{figure}[h]
\includegraphics[width=3in]{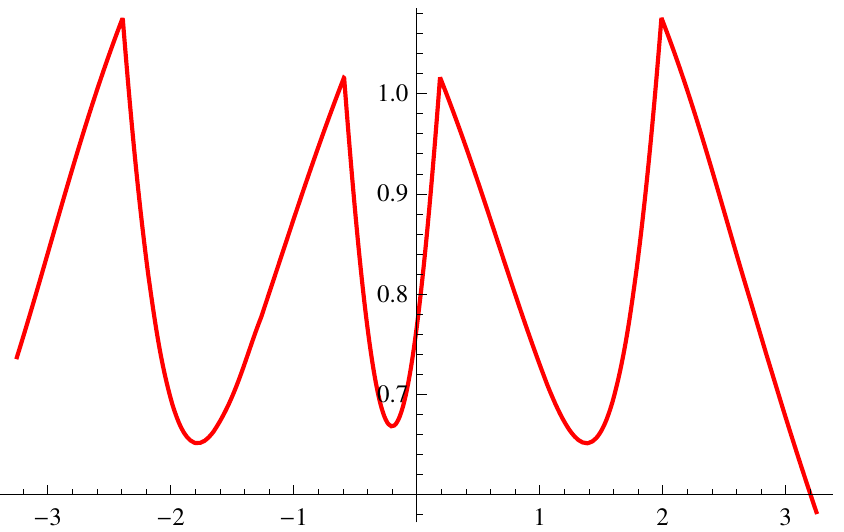}
\caption{The relative magnitude of the magnetic field ($\lvert\vec{B}\rvert$) along the line $x=z=0$ as described by Eq.~(\ref{BAlongYAxis}). Compare this figure to Fig.~\ref{A1A3LocalStats} which shows the degree of alignment along the A1-A3 axis as described in \cite{Hutsemekers}. For this plot we chose $B=1$, $R=1.29$, $d=1.8$ and $s=-0.7$.}
\label{BFieldStrength}
\end{figure}

The influence of the magnetic field itself would be strongest near the loop and fall towards zero away from the loop. While the existence of a background magnetic field does not automatically guarantee an alignment effect, we still expect a higher degree of alignment in regions of space where the magnetic field is stronger. It is interesting to compare the location of high (low) magnetic field strength in Fig.~(\ref{BFieldStrength}) as predicted by our model to the location of high (low) local statistics in Fig.~(\ref{A1A3LocalStats}) as observed in \cite{Hutsemekers}.\footnote{It should be noted that we do not expect the two plots in Fig.~(\ref{BFieldStrength}) and ~(\ref{A1A3LocalStats}) to exactly coincide. Fig.~(\ref{BFieldStrength}) is a plot of the magnitude of the magnetic field along the line $x=z=0$, while the regions A1 and A3 are volumes of space. Furthermore, electroweak cosmic strings are unlikely to be perfectly circular unlike those used in our model. Finally, this issue may also be somewhat alleviated by allowing the two loops to be shifted in the x- and z- directions, in a manner similar to which we included the term $s$ in Eq.~(\ref{BAlongYAxis}). For computational simplicity, however, we chose not to address this as we are more looking for general behavior.}

By allowing the A1-A3 axis to lie along the y-axis, the observed polarization vectors lie entirely in planes of constant y-that is we (approximately) observe only their projection in the x-z plane. Given a magnetic field written as $\vec{B}=B_x\hat{x}+B_y\hat{y}+B_z\hat{z}$, the projected angle of the magnetic field along the $x=z=0$ line (in degrees) can be written as

\be
\theta =\frac{180}{\pi}\arctan \left[\frac{B_x}{B_z}\right]
\label{ProjectedAngleGeneral}
\ee

Because the average angle of polarization is related to the direction of the magnetic field, and the polarization observed is (approximately) entirely in the x-z plane, the direction of polarization can be predicted by plugging Eq.~(\ref{TwoLoops}) into Eq.~(\ref{ProjectedAngleGeneral}).

\be
\theta =\frac{180}{\pi}\arctan \left[\frac{\left(y+s+d\right)\exp\left[-\lvert\lvert y+s+d\rvert -R\rvert\right]}{\left(y+s\right)\exp\left[-\lvert\lvert y+s\rvert -R\rvert\right]}\right]+b
\label{ProjectedAngleSpecific}
\ee

\begin{figure}[h]
\centering
\includegraphics[width=3.4in]{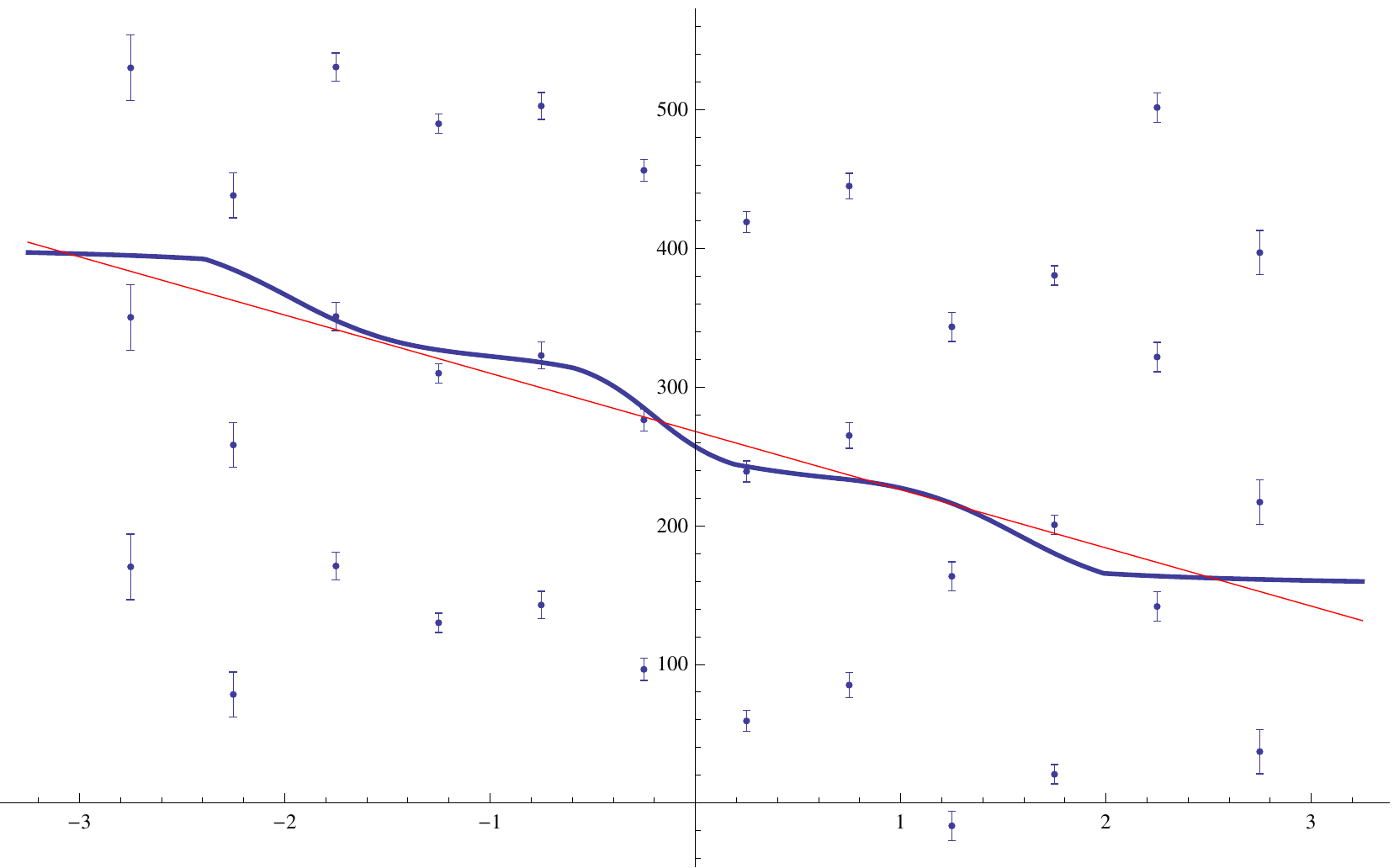}
\caption{Polarization angle vectorally averaged for 183 quasars along the A1-A3 axis from \cite{Hutsemekers}. The objects are divided into bins of size $\Delta z=0.5$. Error bars show the $68 \% $ angular confidence interval as described in \cite{BootstrapConfidenceIntervalEstimation}. The straight red line shown is the linear best fit line from \cite{Hutsemekers} given by $\bar\theta = 268 - 42z$. The blue curved line is our best fit as described by Eq.~(\ref{ProjectedAngleSpecific}). Again, each data point is replicated three times: at $\bar\theta$, $\bar\theta +180$ and $\bar\theta +360$.}
\label{A1A3ErrorBarsAndBestFitLines}
\end{figure}

Fig.~(\ref{A1A3ErrorBarsAndBestFitLines}) shows the vectorally averaged polarization angle as a function of redshift. Because linear polarization angles repeat after a rotation through $180^{o}$, each data point is replicated three times. The term $b$ in Eq.~(\ref{ProjectedAngleSpecific}) is included to allow for an overall shift in angle. Minimizing the chi-square value, we found the parameters in Eq.~(\ref{ProjectedAngleSpecific}) ($s, d, R$ and $b$) best fit by $s=-0.70$, $d=1.80$, $R=1.29$ and $b=324$. Because of the form of the $\arctan$ function, we plotted Fig. (\ref{A1A3ErrorBarsAndBestFitLines}) with $b=324$ for redshift bins centered on $z=-2.75$ through $z=0.25$ and $b=144$ for redshift bins centered on $z=0.75$ through $z=2.75$. The authors of \cite{Hutsemekers} propose a linear best fit given by $\bar\theta = 268 - 42z$. This corresponds to a chi-square value of $98.4$. Our model of Eq.~(\ref{ProjectedAngleSpecific}) with values for the parameters given offers a slight improvement with a chi-square value of $84.2$. We also note that in our analysis to fit the data of Fig. (\ref{A1A3ErrorBarsAndBestFitLines}), we assumed two circular loops; but as discussed in sec. (\ref{EffectFluxAlignment}), electroweak strings contain wiggles and need not even be described by perfectly circular loops. Their shape may bear more resemblance to the shape of the strings in Fig. (\ref{LinkedStrings}). This would certainly alter expected average direction of polarization, and could offer a better fit to the data; although independently determining the exact shape of the specific strings in question here may very well be an extremely difficult task.

\section{Conclusion}

Electroweak strings are predicted to exist in the early universe. Although searching for stable strings in the universe today is a difficult endeavor (as many models predict $\sim 1$ will exist in a given horizon volume), the potential to observe the imprint left over from an electroweak string remains an intriguing possibility. Linked strings may leave behind lines of magnetic flux imprinted in the universe which could be stretched to cosmological scales by both the expansion of the universe and by the fact that parallel lines of magnetic flux repel. Quasars that form in the vicinity of these magnetic fields are essentially forming in a background magnetic field. The quasars would therefore preferentially form with their axes aligned parallel to the magnetic field. The other effect that synergetically works with this is that the external magnetic field tends to align the quasar axis with itself even if initially the two axis were not aligned.
Other nearby quasars will also be forming in essentially the same average background magnetic field which could explain the observed alignment of quasar polarization vectors. On large enough scales, however, the effects of the two looped magnetic fields would be observable as a rotation of the average direction of quasar polarization vectors.

The agreement between our theoretical model and the observational data is very good. In particular we were able to explain the rotation of the polarization angle with the redshift, a feature which is not easily accommodated in simple adhoc models. Our model gives clear predictions that can be tested once a greater sample size of quasar polarization data is available, since we predict an overall trend of quasar polarization vector behavior based on the model given by Eq.~(\ref{TwoLoops}). The A1-A3 axis seems to lie somewhat along the line connecting the NGP and SGP. We would expect other quasar polarization angles in this region to follow the same pattern as observed in Fig. (\ref{A1A3ErrorBarsAndBestFitLines}). Alternately, we may also look for quasar polarization angles away from the A1-A3 axis that still follow the pattern as predicted by Eq.~(\ref{TwoLoops}). This observation would likely be somewhat more difficult, as this would require observing a large sample of objects through the galactic disk. Another interesting test of our model would be to look for the systematic effects such a magnetic field configuration would have on CMB photons such as Faraday rotation.

We emphasis that our explanation of the observed large scale alignment of quasars' polarization angles is based on conventional cosmology and minimal standard model, without invoking any exotic physics or non-standard cosmology. In particular, the formation and subsequent decay of the electroweak strings, and their seeding of the primordial magnetic fields can not be avoided. In this paper we just inked this fact with the large scale alignment of quasars' polarization angles.

\begin{acknowledgments}
%\section{Acknowledgments}
The authors thank D. Dai for the immense help and J. Shea for useful discussion. The authors would also like to express gratitude to N. Kaloper who suggested the idea of explaining quasars' alignment with primordial magnetic fields.  D.S. acknowledges the financial support from NSF, grant number PHY-0914893.

\end{acknowledgments}

\end{document}